\documentclass{aa}

\usepackage{natbib}
\bibpunct{(}{)}{;}{a}{}{,}

\usepackage{txfonts}

\usepackage[dvips]{graphicx}
\usepackage{subfigure}

\begin{document}

\authorrunning{R.~Da~Silva et al.}

\title{Elodie metallicity-biased search for transiting Hot Jupiters
\thanks{Based on radial velocities collected with the ELODIE
spectrograph on the 193-cm telescope and photometric data on the
120-cm telescope, both at the Observatoire de Haute Provence, France}}

\subtitle{I. Two Hot Jupiters orbiting the slightly evolved stars
  \object{HD\,118203} and \object{HD\,149143}}

\author{R.~Da~Silva\inst{1}
\and S.~Udry\inst{1}
\and F.~Bouchy\inst{2}
\and M.~Mayor\inst{1}
\and C.~Moutou\inst{2}
\and F.~Pont\inst{1}
\and D.~Queloz\inst{1}
\and N.C.~Santos\inst{3,1}
\and D.~S\'egransan\inst{1}
\and S.~Zucker\inst{1,4}}

\offprints{R. Da Silva,\\
\email{Ronaldo.daSilva@obs.unige.ch}}

\institute{
Geneva Observatory, 1290 Sauverny, Switzerland
\and
Observatoire de Marseille, France
\and
Centro de Astronomia e Astrof\'\i sica da Universidade de Lisboa,
Tapada da Ajuda, 1349-018 Lisboa, Portugal 
\and
The Weizmann Institute of Science, PO Box 26, Rehovot 76100, Israel
}

\date{29.08.2005 / 22.09.2005}
%
%
\abstract{We report the discovery of a new planet candidate orbiting
  the subgiant star \object{HD\,118203} with a period of
  $P$\,=\,6.1335\,days.  The best Keplerian solution yields an
  eccentricity $e$\,= \,0.31 and a minimum mass
  $m_2\sin{i}$\,=\,2.1\,M$_{\rm Jup}$ for the planet.  This star has
  been observed with the ELODIE fiber-fed spectrograph as one of the
  targets in our planet-search programme biased toward
  high-metallicity stars, on-going since March 2004 at the
  Haute-Provence Observatory. An analysis of the spectroscopic line
  profiles using line bisectors revealed no correlation between the
  radial velocities and the line-bisector orientations, indicating
  that the periodic radial-velocity signal is best explained by the
  presence of a planet-mass companion.  A linear trend is observed in
  the residuals around the orbital solution that could be explained by
  the presence of a second companion in a longer-period orbit. We also
  present here our orbital solution for another slightly evolved star
  in our {\it metal-rich} sample, \object{HD\,149143}, recently
  proposed to host a 4-d period Hot Jupiter by the N2K consortium.
  Our solution yields a period $P$\,=4.09\,days, a marginally
  significant eccentricity $e$\,=\,0.08 and a planetary minimum mass
  of 1.36\,M$_{\rm Jup}$.  We checked that the shape of the spectral
  lines does not vary for this star as well.
  
}

\maketitle
%
%
\section{Introduction}

Stars hosting planets are significantly metal rich in comparison to
field stars in the solar neighbourhood
\citep{Gonzalez1997,Santosetal2001,Santosetal2003,Santosetal2005,FischerValenti2005}.
These authors have also shown that the probability of hosting a giant
planet is a strongly rising function of the star metal content.
According to their estimate, we can expect that up to 25\,-\,30\% of
the more metal-rich stars ([Fe/H]\,$>$\,0.2\,-\,0.3) host a giant
planet.

On the basis of this argument, a new survey was started in March 2004
at the Haute-Provence Observatory with the high-precision ELODIE
fiber-fed echelle spectrograph. The main idea of this new programme is
to bias our target sample towards high-metallicity stars, much more
likely to host planets. This will strongly increase our probability of
finding new planets in a sample of yet non-observed stars.  The survey
uses the cross-correlation technique for the radial-velocity and
metallicity estimates.  The programme mainly targets giant planets
with short periods (Hot Jupiters).  They are the ideal candidates in
the search for photometric transits.  From this survey, we present
here a new short-period planet candidate ($P$\,=\,6.1335\,d) with an
eccentric orbit around the star \object{HD\,118203}.

A similar planet-search programme was simultaneously started by the
N2K consortium \citep{Fischeretal2004} aiming at the detection of
short-period planets orbiting metal-rich stars.  \citet[][ also in
Fischer et al. 2005]{Fischer2005} recently announced two new
Hot-Jupiter detections around \object{HD\,149143} and HD\,109749.
\object{HD\,149143} is amongst the stars already observed in our
sample and we present here our orbital description of the system.

Together, the N2K and ELODIE metallicity-biased planet-search
programmes have detected five new Hot Jupiters in less than one year.
One of them, HD\,149026, is transiting in front of its parent star
\citep{Satoetal2005} allowing for the determination of the planet
radius and mean density. The planet is found to have an unexpectedly
large core. This result clearly illustrates the importance of such
programmes for our understanding of planet interiors. However, when
examining possible statistical trends between orbital and stellar
parameters to derive constraints for planet formation models, we have
to keep in mind the built-in bias of this subsample of exoplanets. In
particular these planets must be removed when considering correlations
with the star metallicity.

The sample selection and observations of the new ELODIE programme are
described in Sect.~\ref{data}. Stellar parameters, radial-velocity
measurements of \object{HD\,118203} as well as the orbital solution
derived for the new Hot Jupiter candidate are presented in
Sect.~\ref{hd118203}.  This section also provides information on
stellar activity and in particular the results of the bisector
analysis for the star.  Section\,\ref{SectHD149143} reports our
results for \object{HD\,149143} and conclusions are presented in
Sect.\,\ref{conc}.

%
%
\section{Sample and observational strategy}
\label{data}
The star sample was selected from the HIPPARCOS catalogue
\citep{ESA1997}.  Initially, the bright northern dwarf stars
(V\,$<$\,8.5) of colours compatible with spectral types ranging from
F8 to M0 (0.45\,$<$\,$B-V$\,$<$\,1.4) were considered. Then:

i) Evolved stars were removed from the list, using a criteria of
proximity to the main sequence in the HR diagram (2\,mag).

ii) All stars already members of the known main programmes of search
for extra-solar planets in the northern hemisphere were removed.

iii) We eliminated stars known as binaries from previous CORAVEL
measurements.

iv) We also eliminated stars with high rotational velocities in order
to avoid active stars in the sample. In practice, stars with
$v\sin{i}$\,$>$\,5\,kms$^{-1}$ were excluded from the sample.

\noindent
The final selection comprises 1061 stars. They lie at distances up to
100\,pc from the Sun.

Observations were conducted with the high-precision ELODIE fiber-fed
echelle spectrograph \citep{Baranneetal1996} mounted on the 1.93-m
telescope at the Haute-Provence Observatory (France). The spectra have
a resolution ($\lambda/\Delta\lambda$) of about 42,000. Typical
signal-to-noise ratios obtained in 20~minute exposures range from
$\sim$20 to 100 for the programme stars, corresponding to photon-noise
errors between 5 and 20~ms$^{-1}$ on individual measurements. The data
reduction is performed on-line during the observations by the
automatic reduction software \citep[see][ for a detailed
description]{Baranneetal1996}.

After a single radial-velocity measurement, it is possible to obtain a
very good estimate of the star metallicity by measuring the surface of
the cross correlation function (CCF) of the ELODIE spectra
\citep{Santosetal2002,Naef2003}. The typical uncertainty of the
resulting metallicities is $\sim$0.05 dex compared to values obtained
from a high-resolution spectroscopic analysis. The expected percentage
of $>$\,10-30\,\% of giant planets orbiting stars with
[Fe/H]\,$\geq$\,0.10
\citep{Santosetal2001,Santosetal2004,FischerValenti2005} should lead
to the discovery of a few tens of new planets in our programme, about
ten of them in short-period orbits.
%
%
\begin{figure}[t!]
  \centering
  \resizebox{0.95\hsize}{!}{\includegraphics{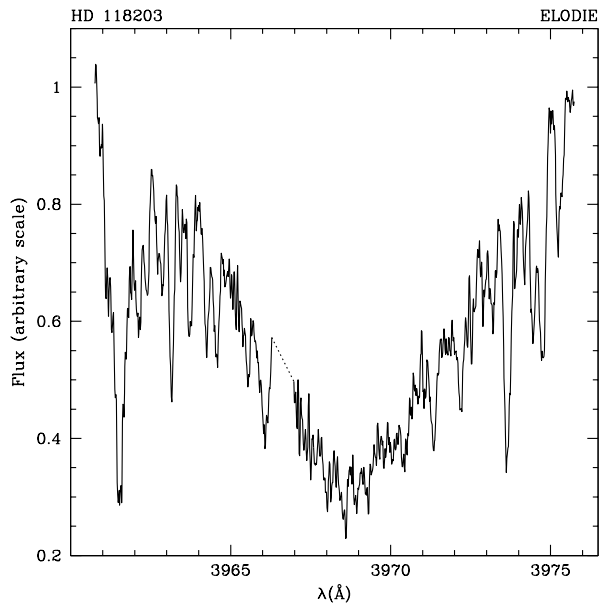}}
  \caption{$\lambda 3968.5$\,\AA~\ion{Ca}{II}~H absorption line region
    of the summed ELODIE spectra for \object{HD\,118203}. No clear
    emission feature is observed. For clarity, spectral features due
    to pollution by the thorium spectrum have been removed (e.g.
    around $\lambda 3966.5$\,\AA).}
  \label{HD118203_Ca}
\end{figure}

\section{A short-period planet orbiting \object{HD\,118203}}
\label{hd118203}

\subsection{Stellar characteristics of \object{HD\,118203}}
  
\object{HD\,118203} (HIP\,66192) is listed in the HIPPARCOS catalogue
\citep{ESA1997} as a K0 dwarf in the northern hemisphere, with a
visual magnitude V\,=\,8.05, a colour index $B-V$\,=\,0.699, and an
astrometric parallax $\pi$\,=\,11.29\,$\pm$\,0.82\,mas, setting a
distance of 88.6\,pc from the Sun. The corresponding absolute
magnitude $M_V$\,=\,3.31 is too high for a K0 dwarf, indicating that
the star is already slightly evolved and is in a subgiant stage.  This
is confirmed by the astrometric surface gravity $\log{g}$\,=\,3.87~dex
estimated from the effective temperature and stellar mass (see below)
and Hipparcos precise parallax \citep{Santosetal2004}.
  
From calibrations of the width and the surface of the ELODIE
cross-correlation functions \citep{Santosetal2002,Naef2003}, we have
estimated a projected rotation velocity $v{\sin
  i}$\,=\,4.7\,kms$^{-1}$ and a metallicity [Fe/H]\,=\,0.10 for the
star.  Comparing these {\it observable} stellar characteristics with
the stellar evolution models of \citet{Girardietal2002}, we can infer
the following intrinsic properties for \object{HD\,118203}: a coarse
estimate of the effective temperature $T_{\rm
  eff}$\,=\,5600\,$\pm$\,150\,K, a mass
$M_\star$\,=\,1.23\,$\pm$\,0.03\,M$_\odot$ and an age of
4.6\,$\pm$\,0.8\,Gyr.  A similar value of the effective temperature
$T_{\rm eff}$\,=\,5695\,$\pm$\,50\,K is also derived from the
calibration in \citet{Santosetal2004} using the color index and
metallicity of the star.  The stellar parameters are gathered in
Table\,\ref{stellar_par}.

A check for stellar chromospheric activity was also performed by
looking at the \ion{Ca}{II}~H absorption line in the spectra.
Figure~\ref{HD118203_Ca} shows the co-added ELODIE spectra where the
clear absence of an emission feature indicates a low activity level,
as expected for slightly evolved subgiants \citep{Wright2004}.  The
activity-induced radial-velocity jitter is thus also expected to be
low for this moderately rotating star.

\begin{table}[t!]
\centering
  \caption[]{Observed and estimated parameters for \object{HD\,118203}.
(See text for references on the quoted values.)}
  \label{stellar_par}
\begin{tabular}{lcl}
\hline
\hline
\noalign{\smallskip}
Spectral Type         & K0               &             \\
$V$                   & 8.05             &             \\
$B-V$                 & 0.699            &             \\
$\pi$                 & 11.29 $\pm$ 0.82 & [mas]       \\
$M_V$                 & 3.31             &             \\
$T_{\rm eff}$         & 5600 $\pm$ 150   & [K]         \\
$M_\star$             & 1.23 $\pm$ 0.03  & M$_{\odot}$ \\
{\it age}             & 4.6  $\pm$ 0.8   & Gyr         \\
$\log{g}$              & -3.87           & cgs         \\ 
$[$Fe/H$]$            & 0.10 $\pm$ 0.05  &             \\
$v{\sin i}$           & 4.7              & [kms$^{-1}$]      \\
\noalign{\smallskip}
\hline
\end{tabular}
\end{table}
%
%
\subsection{Orbital solution for \object{HD\,118203}\,b}
\label{orb_solution}

ELODIE observations of \object{HD\,118203} were conducted from May
2004 (JD\,=\,2\,453\,151) to July 2005 (JD\,= \,2\,453\,553),
stimulated by the difference in radial velocity found among the first
three observations. A set of 43 precise radial-velocity measurements
were then gathered. They are provided in Table\,\ref{table_rv1}.

A clear 6.1-d periodic variation is seen in this data set.  The
residuals around a Keplerian fit with this period are however
unexpectedly large ($>$\,40\,ms$^{-1}$) and shows a clear additional
radial-velocity drift as a function of the Julian date. A simultaneous
Keplerian + linear-drift adjustment then provides a very satisfactory
solution for the system with a period
$P$\,=\,6.1335\,$\pm$\,0.0006~days and an eccentricity
$e$\,=\,0.309\,$\pm$\,0.014.  The slope of the linear drift is found
to be 49.7\,$\pm$\,5.7\,m${\rm s}^{-1}{\rm yr}^{-1}$.  It is most
probably accounted for by the presence of a second companion in the
system, on a longer-period orbit.  This could explain the
significantly non-zero value found for the eccentricity at such a
short period\footnote{The period is however close to (higher than?)
  the observed circularisation period determined from the sample of
  known exoplanets around 6 days \citep{Halbwachsetal2004}.}.

\begin{figure}[t]
  \centering
    \centering
    \resizebox{0.95\hsize}{!}{\includegraphics{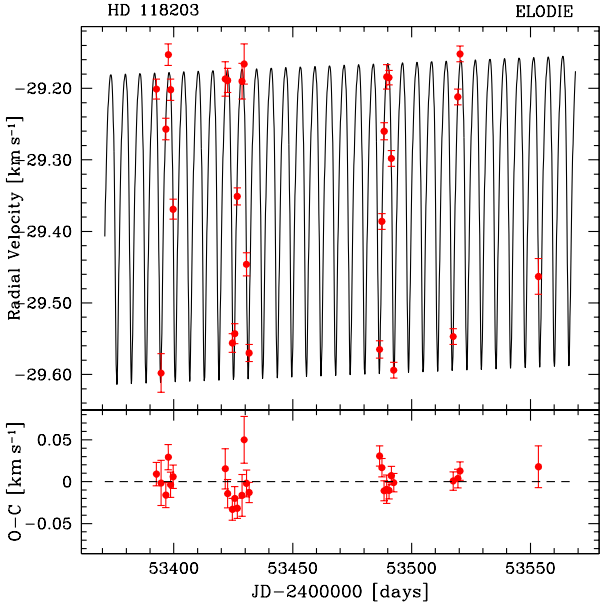}}
    \caption{Top: last 5-months radial-velocity measurements of
      \object{HD\,118203} superimposed on the best Keplerian + linear
      drift solution indicative of a second longer-period companion in
      the system.  Bottom: residuals around the solution. Error bars
      represent the photon-noise errors.}
    \label{hd118203_dvft}
\end{figure}

\begin{figure}[t!]
  \centering
  \resizebox{0.95\hsize}{!}{\includegraphics{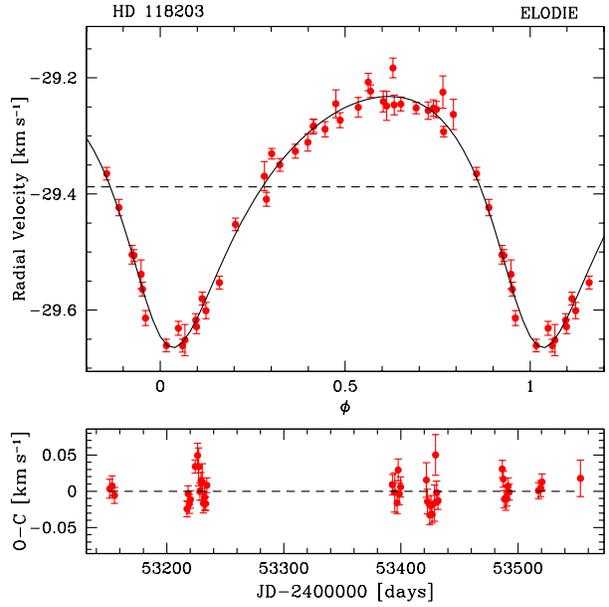}}
  \caption{Top: phase-folded ELODIE radial-velocity measurements for
    the star \object{HD\,118203} after correction for the linear
    radial-velocity drift observed in the data.  Bottom: corresponding
    residuals around the solution. Error bars represent the
    photon-noise errors.}
    \label{hd118203_ph}
\end{figure}

A plot with the last-season radial-velocity measurements of
\object{HD\,118203} is shown in Fig.~\ref{hd118203_dvft}, together
with the derived solution.  The residuals around the solution are
displayed as well in the bottom panel of the figure. The phase-folded
radial-velocity curve with the complete set of data points corrected
for the observed linear drift is displayed in Fig.~\ref{hd118203_ph}.
The weighted rms around the solution is
$\sigma$(O$-$C)\,=\,18.1\,ms$^{-1}$, slightly larger than the
individual photon-noise errors ($\sim$15\,ms$^{-1}$).  The orbital
elements are listed in Table~\ref{orb_elem} with the inferred
planetary parameters.

\begin{table}[t!]
\caption{ELODIE radial velocities of \object{HD\,118203}. 
All data are relative to the solar system barycentre.
Given uncertainties correspond to photon-noise errors.}
\label{table_rv1}
\centering
\begin{tabular}{c c c}
\hline\hline
\bf JD-2400000 & \bf RV & \bf Uncertainty \\
\bf [days] & \bf [km\,s$^{-1}$] & \bf [km\,s$^{-1}$] \\
\hline
 53151.4302  & $-29.263$  &     0.013 \\
 53153.4235  & $-29.231$  &     0.014 \\
 53155.3946  & $-29.640$  &     0.011 \\
 53217.3409  & $-29.523$  &     0.011 \\
 53218.3429  & $-29.320$  &     0.011 \\
 53219.3450  & $-29.243$  &     0.013 \\
 53220.3498  & $-29.215$  &     0.012 \\
 53224.3411  & $-29.300$  &     0.009 \\
 53226.3539  & $-29.152$  &     0.017 \\
 53227.3547  & $-29.232$  &     0.026 \\
 53228.3326  & $-29.533$  &     0.012 \\
 53229.3231  & $-29.549$  &     0.011 \\
 53230.3545  & $-29.338$  &     0.025 \\
 53231.3597  & $-29.257$  &     0.013 \\
 53232.3270  & $-29.209$  &     0.018 \\
 53233.3267  & $-29.261$  &     0.009 \\
 53234.3264  & $-29.474$  &     0.010 \\
 53392.6777  & $-29.201$  &     0.014 \\
 53394.6422  & $-29.598$  &     0.027 \\
 53396.6787  & $-29.257$  &     0.015 \\
 53397.6810  & $-29.153$  &     0.015 \\
 53398.6756  & $-29.202$  &     0.015 \\
 53399.6799  & $-29.369$  &     0.014 \\
 53421.6764  & $-29.187$  &     0.024 \\
 53422.6439  & $-29.189$  &     0.017 \\
 53424.6543  & $-29.556$  &     0.013 \\
 53425.6554  & $-29.543$  &     0.014 \\
 53426.6586  & $-29.351$  &     0.012 \\
 53428.6526  & $-29.190$  &     0.025 \\
 53429.5865  & $-29.166$  &     0.028 \\
 53430.5650  & $-29.446$  &     0.016 \\
 53431.6340  & $-29.570$  &     0.012 \\
 53486.5332  & $-29.565$  &     0.012 \\
 53487.4809  & $-29.386$  &     0.011 \\
 53488.4736  & $-29.260$  &     0.012 \\
 53489.5168  & $-29.184$  &     0.017 \\
 53490.4763  & $-29.185$  &     0.010 \\
 53491.4789  & $-29.298$  &     0.011 \\
 53492.4688  & $-29.594$  &     0.011 \\
 53517.4910  & $-29.547$  &     0.011 \\
 53519.4450  & $-29.212$  &     0.011 \\
 53520.3896  & $-29.152$  &     0.011 \\
 53553.3845  & $-29.463$  &     0.025 \\
\hline
\end{tabular}
\end{table}

%
\subsection{Low chromospheric activity for \object{HD\,118203}}
\label{bisec}
Physical events in the stellar atmosphere, like the presence of spots
on the stellar surface, can change the observed spectral-line profiles
and induce a transient periodic radial-velocity signal similar to the
one expected from the presence of a planet. The bisector analysis is
one of the best tools to discriminate between radial-velocity
variations due to changes in the spectral-line shapes and variations
due to the real orbital motion of the star \citep[see][ for a
description of this method]{Quelozetal2001}.

\begin{figure}[h!]
  \centering
  \resizebox{0.74\hsize}{!}{\includegraphics{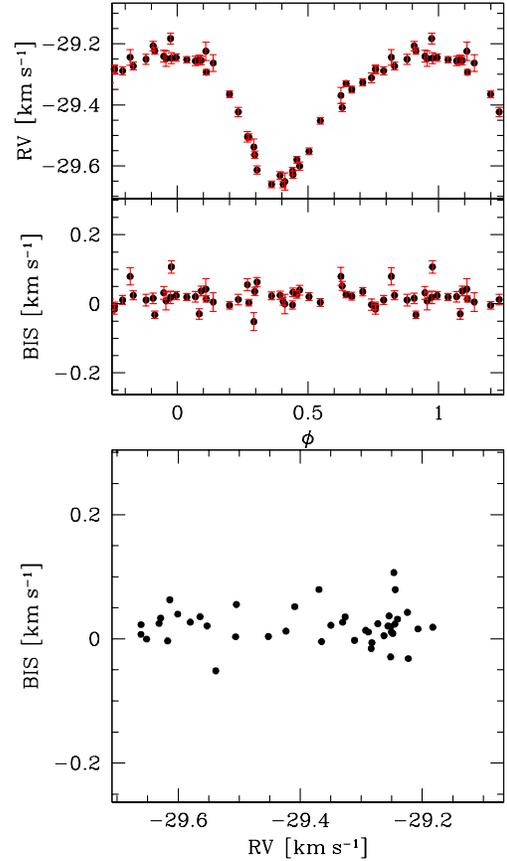}}
  \caption{Radial velocities (RV, upper panel) and inverse bisector
    slope (BIS, middle panel) phased with the orbital period
    $P$\,=\,6.1335\,d for \object{HD\,118203}. The independence of the
    two quantities is shown by the radial velocity vs. BIS plot
    (bottom panel).}
  \label{bisec_fig}
\end{figure}

The bisector inverse slope (BIS value) computed from the
\object{HD\,118203} spectra are plotted in Fig.~\ref{bisec_fig} in
comparison to the corresponding radial-velocities.  If the line-shape
and the radial-velocity variations share the same origin, the BIS and
radial velocities are expected to be correlated.  This is not the
case.  Furthermore, no coherent signal is observed when the BIS are
phased with the orbital period of 6.1335\,days.

%
%
\begin{table}[h]
\centering
  \caption[]{ELODIE best Keplerian orbital solution obtained 
   for \object{HD\,118203} as well as the inferred planetary
   parameters. For eccentric orbits
  $T$ is defined as the time of the peri-astron passage.}
  \label{orb_elem}
\begin{tabular}{l r@{ }l l}
\hline
\hline
\noalign{\smallskip}
$P$                  & 6.1335    & $\pm$ 0.0006 & [days]               \\
$T$                  & 53394.23  & $\pm$ 0.03   & [JD $-$ 2\,400\,000] \\
$e$                  & 0.309     & $\pm$ 0.014  &                      \\
$V$                  & $-$29.387 & $\pm$ 0.006  & [km ${\rm s}^{-1}$]  \\
$\omega$             & 155.7     & $\pm$ 2.4    & [deg]                \\
$K$                  & 217       & $\pm$ 3      & [m ${\rm s}^{-1}$]   \\
$N_{\rm meas}$       & 43        &              &                      \\
$\sigma$(O$-$C)      & 18.1      &              & [m ${\rm s}^{-1}$]   \\
Linear drift         & 49.7      & $\pm$ 5.7    &
                                      [m ${\rm s}^{-1}{\rm yr}^{-1}$]  \\
\noalign{\smallskip}
\hline
\noalign{\smallskip}
$a_{\rm 1} {\sin i}$ & 11.61     &              & [$10^{-5}$ AU]       \\
$f(m)$               & 5.55      &              & [$10^{-9} M_\odot$]  \\
$m_2 {\sin i}$       & 2.13      &              & [$M_{\rm Jup}$]      \\
$a$                  & 0.07      &              & [AU]                 \\
\noalign{\smallskip}
\hline
\end{tabular}
\end{table}

Moreover, data gathered during the still negative search for a
photometric transit (presented elsewhere) show that the star is
constant at a 0.0047~mag level and thus further support its
low-activity level.  Consequently, the explanation of the observed
radial-velocity variations due to stellar activity can be discarded.


%
\section{A Hot Jupiter around \object{HD\,149143}}
\label{SectHD149143}

During the preparation of this publication, another new Hot Jupiter
candidate was identified in our programme, orbiting the star
\object{HD\,149143}. While gathering more points to characterize the
system, we learned that the planet had just been announced by the N2K
consortium \citep{Fischer2005,Fischeretal2005}.  We thus present here
our ELODIE solution for the system, relying on the N2K photometric
measurements for the transit non-detection.

\begin{figure}[t]
  \centering
  \resizebox{0.95\hsize}{!}{\includegraphics{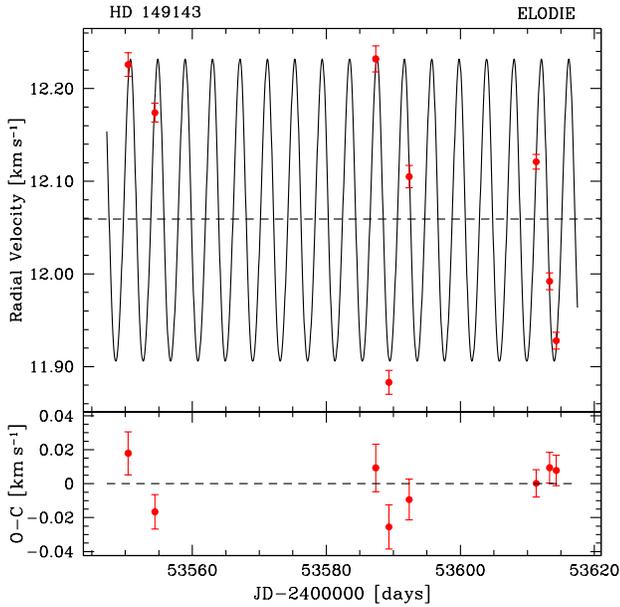}}
  \caption{ELODIE radial velocities of \object{HD\,149143}
    superimposed on the derived Keplerian quasi-circular model.
    Residuals around the solution are displayed in the bottom panel.
    Error bars represent the photon-noise errors.}
  \label{hd149143_dvft}
\end{figure}

\begin{table}[t]
\centering
  \caption[]{Observed and estimated parameters for \object{HD\,149143}.
(See text for references on the quoted values.)}
  \label{table_star2}
\begin{tabular}{lcl}
\hline
\hline
\noalign{\smallskip}
Spectral Type         & G0               &             \\
$V$                   & 7.9              &             \\
$B-V$                 & 0.68             &             \\
$\pi$                 & 15.75 $\pm$ 1.07 & [mas]       \\
$M_V$                 & 3.88              &             \\
$T_{\rm eff}$         & 5730 $\pm$ 150   & [K]         \\
$M_\star$             & 1.1 $\pm$ 0.1    & M$_{\odot}$ \\
{\it age}             & 7.6 $\pm$ 1.2    & Gyr         \\
$\log{g}$             & 4.10             & cgs         \\
$[$Fe/H$]$            & 0.20 $\pm$ 0.05  &             \\
$v{\sin i}$           & 3.9              & [kms$^{-1}$]      \\
\noalign{\smallskip}
\hline
\end{tabular}
\end{table}

\subsection{\object{HD\,149143}: stellar characteristics}

\object{HD\,149143} is listed in the Hipparcos catalogue (HIP\,81022)
with a G0 spectral type, a visual magnitude $V$\,=\,7.90, and a colour
index $B$\,-\,$V$\,=\,0.68.  From calibrations of the width and the
surface of the ELODIE cross-correlation functions we have estimated a
projected rotation velocity $v{\sin i}$\,=\,3.9\,kms$^{-1}$ and a
metallicity [Fe/H]\,=\,0.2.

\begin{figure}[h!]
  \centering
  \resizebox{0.9\hsize}{!}{\includegraphics{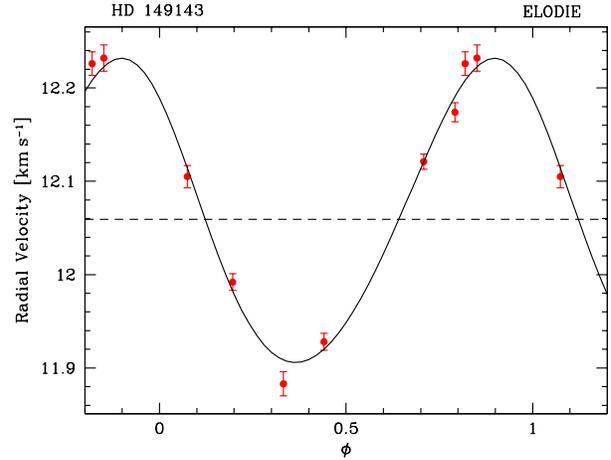}}
  \caption{Phase-folded ELODIE radial velocities of \object{HD\,149143}
    superimposed on the derived quasi-circular Keplerian solution.
    Residuals around the solution are displayed in the bottom panel.
    Error bars represent the photon-noise errors.}
  \label{hd149143_phas}
\end{figure}

The measured Hipparcos parallax of \object{HD\,149143} is
$\pi$\,=\,15.75\,$\pm$\,1.07\,mas setting the star 63.5\,pc away from
the Sun. The inferred absolute magnitude is then $M_V$\,=\,3.88, also
slightly overluminous for a metallic G0 dwarf. This is confirmed by
the astrometric estimate of $\log{g}$\,=\,4.10~dex.  From the models
of \citet{Girardietal2002} we obtain an effective temperature $T_{\rm
  eff}$\,=\,5730\,$\pm$\,150\,K, a primary mass
$M_1$\,=\,1.1\,$\pm$\,0.1\,M$_\odot$ and an age of
7.6\,$\pm$\,1.2\,Gyr. A calibration of the stellar effective
temperature based on the stellar colour and metallicity \citep[][ only
slightly sensitive to the gravity of the star]{Santosetal2004} leads
then to $T_{\rm eff}$\,=\,5790\,$\pm$\,50\,K, compatible with the
value given above. The stellar parameters are given in
Table\,\ref{table_star2}.

\subsection{Orbital solution for \object{HD\,149143}\,b}

Although the number of ELODIE observations is small for this star
(Table\,\ref{table_rv2}), thanks to the large amplitude of the
radial-velocity variation ($K$\,=\,163\,ms$^{-1}$), we can easily
derive a quasi-circular Keplerian solution ($e$\,=\,0.08\,$\pm$\,0.04)
with a period of 4.088~days, in agreement with the \citet{Fischer2005}
and \citet{Fischeretal2005} announcement.  The eccentricity of the
orbit is small and marginally significant, we thus also provide the
corresponding circular solution fixing e=0 (Table\,\ref{orb_elem2}).
The derived orbital elements coupled with the above estimate of the
primary mass of 1.1\,M$_{\odot}$ lead to a minimum mass
$m_2\sin{i}$\,=\,1.36\,M$_{\rm Jup}$ and a separation of 0.052\,AU for
the planetary companion.

Figures\,\ref{hd149143_dvft} and \ref{hd149143_phas} present the
temporal and phased ELODIE radial-velocity measurements superimposed
on the derived Keplerian model as well as the residuals around the
solution. The inferred orbital and planetary parameters are given in
Table\,\ref{orb_elem2}.

As for the case of \object{HD\,118203}, the bisector inverse slope of
the cross-correlation function has been calculated. No correlation is
found between these bisector slopes and the radial velocities,
excluding activity-induced variations of the shape of the spectral
lines as the source of the radial-velocity variations.

%
\begin{table}
\caption{ELODIE radial velocities of \object{HD\,149143}. 
All data are relative to the solar system barycentre.}
\label{table_rv2}
\centering
\begin{tabular}{c c c}
\hline\hline
\bf JD-2400000 & \bf RV & \bf Uncertainty \\
\bf [days] & \bf [km\,s$^{-1}$] & \bf [km\,s$^{-1}$] \\
\hline
 53550.4467     & 12.226  & 0.013 \\
 53554.4224     & 12.174  & 0.010 \\
 53587.3709     & 12.232  & 0.014 \\
 53589.3393     & 11.880  & 0.013 \\
 53592.3736     & 12.102  & 0.012 \\
 53611.3172     & 12.121  & 0.010 \\
 53613.3130     & 11.992  & 0.010 \\
 53614.3141     & 11.928  & 0.010 \\
\hline
\end{tabular}
\end{table}
\begin{table}
\centering
  \caption[]{ELODIE derived Keplerian orbital solution obtained 
  for \object{HD\,149143} as well as the inferred planetary 
  parameters. Both the quasi-circular and circular solutions are given 
  because the derived eccentricity is only  marginally significant.
  For the eccentric orbit $T$ is defined as the time of the 
  peri-astron passage whereas for
  the circular orbit $T$ indicates the maximum of radial velocities.}
  \label{orb_elem2}
\begin{tabular}{l r@{ }l r@{ }l l}
\hline
\hline
\noalign{\smallskip}                                                       
 & \multicolumn{2}{c}{$e$ free} & \multicolumn{2}{c}{$e=0$} & \\
\noalign{\smallskip}                                                       
\hline
$P$             &4.088   &$\pm$ 0.006 &4.089   &$\pm$ 0.006 &[days]               \\
$T$             &53588.0 &$\pm$ 0.4   &53587.4 &$\pm$ 1.0   &[JD\,$-$\,2\,400\,000] \\
$e$             &0.08    &$\pm$ 0.04  &0.      &            &                      \\
$V$             &12.059  &$\pm$ 0.004 &12.056  &$\pm$ 0.003 &[kms$^{-1}$]  \\
$\omega$        &42      &$\pm$ 35    &0.      &            &[deg]                \\
$K$             &163     &$\pm$ 8     &156     &$\pm$ 6     &[ms$^{-1}$]   \\
$\sigma$(O$-$C) &13.3    &            &16.0    &            &[ms$^{-1}$]   \\
\noalign{\smallskip}                                                       
\hline                                                                     
\noalign{\smallskip}                                                       
$a_{\rm 1} {\sin i}$ & 6.105   &  & 5.847     &  & [$10^{-5}$ AU]       \\
$f(m)$               & 1.82    &  & 1.59      &  & [$10^{-9} M_\odot$]  \\
$m_2 {\sin i}$       & 1.36    &  & 1.30      &  & [$M_{\rm Jup}$]      \\
$a$                  & 0.052   &  & 0.052     &  & [AU]                 \\
\noalign{\smallskip}
\hline
\end{tabular}
\end{table}

\section{Summary and concluding remarks}
\label{conc}
  
We have presented the characteristics of a new planet candidate in
orbit around the subgiant star \object{HD\,118203}, detected by the
new ELODIE planet-search programme biased towards metal-rich stars.
The planet is in a rather eccentric orbit ($e$\,=\,0.31), with a
period of $P$\,=\,6.1335~days, and is close to its parent star
($a$\,=\,0.06\,AU).

An additional trend of the radial-velocity measurements increasing as
a function of Julian date with a slope of 49.7~ms$^{-1}{\rm yr}^{-1}$
is observed, suggesting the presence of a second companion around the
main star on a longer-period orbit. This additional companion could
explain the somewhat high eccentricity of the orbit of this new planet
at short period.

We have also reported the ELODIE solution for another Hot Jupiter in
our programme, \object{HD\,149143}, recently announced by the
similar-goal N2K project. The derived best Keplerian quasi-circular
solution presents a period of 4.09\,days and the inferred planetary
mass is 1.36\,M$_{\rm Jup}$.

By selection these planets are orbiting metal-rich stars. They
increase to five the number of Hot Jupiters detected in less than
one~year by dedicated metallicity-biased programmes. One of these
candidates, HD\,149026 \citep{Satoetal2005}, transits in front of its
parent star and thus allows the determination of its radius and mean
density when combining photometric and radial-velocity measurements.
This demonstrates the efficiency of such approaches to find candidates
suitable for constraining planet-interior models.  However, the
built-in biases of the sample have to be kept in mind when examining
possible statistical relations between the star metallicity and other
orbital or stellar parameters.

\begin{acknowledgements}
  We thank the Swiss National Science Foundation (FNSRS) and the
  Geneva University for their continuous support to our planet-search
  programmes. We also thank the Haute-Provence Observatory for the
  granted telescope time. Support from Funda\c{c}\~ao para a Ci\^encia
  e a Tecnologia (Portugal) to N.C.S.  in the form of a scholarship
  (reference SFRH/BPD/8116/2002) and a grant (reference
  POCI/CTE-AST/56453/2004) and support from Coordena\c c\~ao de
  Aperfei\c coamento de Pessoal de N\'\i vel Superior (CAPES - Brazil)
  to R.D.S. in the form of a scholarship are gratefully acknowledged
  as well.
\end{acknowledgements}

\bibliographystyle{aa}
\bibliography{aahd118203}

\end{document}